\documentclass[a4paper,11pt]{article}
\usepackage{jheppub}

\usepackage{amssymb}
\usepackage{amsmath}
\usepackage{amscd}
\usepackage{latexsym}
\usepackage{epsfig}
\usepackage{color}
\usepackage{fancyhdr}
\usepackage[bf,footnotesize]{caption}
\usepackage{graphicx}
\usepackage{booktabs}

\hypersetup{%
  bookmarksnumbered=true,
  pdftitle = {},
  pdfsubject = {},
  pdfauthor = {U.~Wenger},
  pdfkeywords = {}
}

\graphicspath{{Figures/}}

\def\Dslash{\hbox{D}\kern-0.6em\raise0.15ex\hbox{/}} 

\def\R{{\mathbb R}}                     

\def\onehalf{\frac{1}{2}}               

\def\bi{\begin{itemize}}
\def\ei{\end{itemize}}
\def\be{\begin{equation}}
\def\ee{\end{equation}}

\def\DW{D_\text{W}}
\def\D5{D_5}

\date{\empty}

\title{Spectral properties of the Wilson Dirac operator\\
 in the $\epsilon$-regime}

\author{Albert Deuzeman,}
\author{Urs Wenger}
\author{and Ja\"ir Wuilloud}

\affiliation{
Albert Einstein Center for Fundamental Physics,\\
Institute for Theoretical Physics,\\
Sidlerstrasse 5, CH--3012 Bern, Switzerland.}

\emailAdd{deuzeman@itp.unibe.ch}
\emailAdd{wenger@itp.unibe.ch}
\emailAdd{jair@itp.unibe.ch}

\abstract{
  We investigate the spectral properties of the Wilson Dirac operator
  in quenched QCD in the microscopic regime.  We distinguish the
  topological sectors using the index as determined by the Wilson flow
  method. Consequently, the distributions of the low-lying eigenvalues
  of the Wilson Dirac operator can be compared in each of the
  topological sectors to predictions from random matrix theory applied
  to the $\epsilon$-regime of chiral perturbation theory. We find rather
  good agreement for volumes as small as $(1.5 \, \text{fm})^4$ and lattice
  spacings as coarse as $0.1\, \text{fm}$, and demonstrate that it is indeed
  possible to extract low-energy constants for Wilson fermions from
  the spectral properties of the Wilson Dirac operator.}


\arxivnumber{1110.4002}

\begin{document}

\maketitle
\flushbottom

\section{Introduction}
The spontaneous breaking of chiral symmetry in Quantum Chromodynamics
(QCD) is an important non-perturbative feature that determines the
low-energy behaviour of the theory in a crucial way. In chiral
perturbation theory, an effective description of QCD at low energies,
the effect of spontaneous chiral symmetry breaking is encoded in the
quark condensate $\Sigma$ and the Goldstone boson fields as the
relevant degrees of freedom
\cite{Weinberg:1968de,Gasser:1983yg,Gasser:1984gg}. The effective
theory then describes their dynamics while being strongly constrained
by the requirements of chiral symmetry. In a specific regime of the
theory, the so-called $\epsilon$-regime
\cite{Gasser:1987ah,Gasser:1987zq,Hasenfratz:1989pk,Hansen:1990un}
where the mass $m_\pi$ of the Goldstone pion is small compared to the linear
extent $L$ of the space-time volume $V=L^4$, i.e.~$m_\pi L \ll 1$, the
finite volume partition function is dominated by the zero-momentum
mode of the pion field. In this situation, the topology of the gauge
fields becomes important and as a consequence the partition function
naturally separates into sectors of fixed topology
\cite{Leutwyler:1992yt}.

Intriguingly, in the $\epsilon$-regime there exists
another effective description of QCD which is based on random matrices
\cite{Shuryak:1992pi,Verbaarschot:1993pm,Verbaarschot:1994qf}. In
Random Matrix Theory (RMT) one models the fluctuations and
correlations of the energy levels of the Hamilton operator of a
physical system by considering the eigenvalues of matrices with random
entries, but obeying the same global symmetries.  For describing the
Euclidean QCD partition function, for example, one may construct a
chiral random matrix model for the massless QCD Dirac operator. It
turns out that in the microscopic regime this leads to the proper
accumulation of eigenvalues at the origin, a characteristic feature
of spontaneous chiral symmetry breaking.  By now, one has a complete
understanding of how and why chiral RMT represents an exact limit of
the light quark partition function of QCD in the $\epsilon$-regime and
a precise mapping between QCD, standard chiral perturbation theory and
chiral RMT is established (see
e.g.~\cite{Verbaarschot:2000dy,Damgaard:2011gc} for a review and
further references). As a consequence, RMT provides an interesting
alternative approach for studying the universal properties of
low-energy QCD.

What is common to both effective theories is the fact that their
intrinsic parameters, the effective low-energy constants, need to be
determined externally, either from phenomenology, or directly from QCD
using calculations from first principles, e.g.~lattice QCD. In the
past, there have been several efforts to extract physical parameters
from numerical simulations in the $\epsilon$-regime, but all of them
have been restricted to the use of lattice Dirac operators which obey
an exact chiral symmetry
\cite{Bietenholz:2003mi,Giusti:2003gf,DeGrand:2006uy}. Recently it has
become evident, however, that simulations in the $\epsilon$-regime
need not exclusively be reserved to these actions, but can also be
done using standard Wilson fermions for which the chiral symmetry is
broken by the discretisation. It has been understood in the context of
Wilson chiral perturbation theory (WChiPT) how to include the effects
of the Wilson lattice discretisation in the standard regime of the
theory \cite{Sharpe:1998xm,Rupak:2002sm,Bar:2003mh} and this has also
been achieved in the $\epsilon$-regime
\cite{Bar:2008th,Shindler:2009ri}. Remarkably, also chiral RMT can be
extended to include the effects of the Wilson lattice discretisation,
leading to Wilson chiral Random Matrix Theory (WRMT)
\cite{Damgaard:2010cz,Akemann:2010em}.

Subsequently, a series of interesting statements have been made about
the spectral density of the Wilson Dirac operator using WChiPT and/or
WRMT for both $N_f=0$ (quenched) and $N_f=1$ QCD
\cite{Damgaard:2010cz,Akemann:2010em,Kieburg:2011uf,Necco:2011vx,
  Necco:2011jz,Akemann:2010zp,Splittorff:2011bj}. One motivation is
that a good description of the Wilson Dirac operator spectrum is most
important for lattice simulations towards the chiral limit at fixed
lattice spacing. In order to perform such simulations, a thorough
understanding of the quark mass, volume and lattice spacing dependence
of the spectrum is crucial for avoiding potential numerical
instabilities.  In that respect, the quenched approximation which we
are studying in this paper can still be useful for $N_f>0$ simulations
of QCD where $m\Sigma V \gg 1$, because in that case the spectrum of
the massless Dirac operator behaves essentially as in quenched QCD.

Since topology plays a crucial role in the $\epsilon$-regime it is
important to have good control over the topological properties of the
system. A natural definition for the topological charge of the gauge
fields is provided by the Atiyah-Singer index theorem which can be
carried over to the lattice for chirally symmetric lattice Dirac
operators \cite{Hasenfratz:1998ri}. One problem with simulating
Wilson fermions in this extreme regime, though, is related to the
fact that the chiral index is not well defined for the Wilson Dirac
operator.

However, there exists an exact relation between the number of real
modes of the Wilson Dirac operator and the index as defined from the
overlap operator \cite{Narayanan:1994gw,Neuberger:1997fp} using the
Wilson Dirac operator as its kernel. One way to define a topological
index for the Wilson Dirac operator is then provided by the eigenvalue
flow of the hermitian Wilson Dirac operator, the so-called Wilson flow
method
\cite{Itoh:1987iy,Narayanan:1994gw,Edwards:1998sh,Edwards:1998gk,Edwards:1999ra}.
The definition becomes unambiguous in the continuum limit and may
hence serve as a natural definition of the index. It is an astonishing
result of the WRMT descriptions
\cite{Damgaard:2010cz,Akemann:2010em,Kieburg:2011uf} that they can
actually make definite statements about the hermitian eigenvalue flow
and therefore make a natural connection to the various topological
charge sectors of QCD.

In this paper we investigate the spectral properties of the Wilson
Dirac operator in quenched QCD in the $\epsilon$-regime corresponding
to the microscopic regime of WRMT by means of numerical
simulations. We calculate the low-lying spectrum of the hermitian
Wilson Dirac operator and compare the distributions of the single
eigenvalues to the ones from WRMT. Once the matching between the QCD
data and WRMT is accomplished, we can check the predictive power of
WRMT by scaling and comparing the results at different masses and
volumes. For distinguishing the topological charge sectors we
determine the Wilson flow on each gauge field configuration. As a
by-product, the Wilson flow also yields the distribution of the real
modes of the Wilson Dirac operator and this can again be compared to
the chirality distributions as described by WRMT.

\section{Wilson random matrix theory}
We start by briefly recalling the WRMT that is expected to describe
the spectrum of the Wilson Dirac operator. The Symanzik expansion has
three operators -- $W_6$, $W_7$ and $W_8$ -- that provide the leading
order corrections for WChiPT. Arguments from a large-$N_c$ expansion
indicate that the terms proportional to $W_6$ and $W_7$ are
suppressed~\cite{Kaiser:2000gs} and these terms are often set to zero
in a first approximation.

We consider matrices of the form
\be
\label{eq:DW_RMT}
\tilde D^\nu_\text{W} = 
\left(
\begin{array}{cc}
\tilde a_8 A & i W \\
 i W^\dagger & \tilde a_8 B
\end{array}
\right) \, ,
\ee
where $A=A^\dagger$ and $B=B^\dagger$ are $(n+\nu)\times (n+\nu)$ and
$n\times n$ square matrices, respectively, while $W$ is an arbitrary
complex $(n+\nu)\times n$ rectangular matrix \cite{Hehl:1998pt, Akemann:2010em}. We use tildes to
distinguish the quantities from the analogous ones in the quantum
field theory setup. The block structure of $\tilde D^\nu_\text{W}$ can
be interpreted as a chiral decomposition with additional terms
$\propto \tilde a_8$ mixing the chiral sectors and thereby incorporating
the effects of the operator $W_8$.  Furthermore, the structure
in eq.(\ref{eq:DW_RMT}) guarantees that there are at least $|\nu|$
real eigenvalues. We are interested in the eigenvalues of
the hermitian matrix
\be
\tilde D_5^\nu (\tilde m)  = \tilde \gamma_5 (\tilde D_\text{W}^\nu + \tilde m) 
\ee
where $\tilde \gamma_5 = \text{diag}(1,\ldots,1,-1,\ldots,-1)$ with
$(n+\nu)$ entries of 1 and $n$ entries of $-1$ along the
diagonal.

The matrix elements of $\tilde D^\nu_\text{W}$  are randomly distributed according to the Gaussian
weight
\be
P(A,B,W) = e^{-\frac{N}{4} \text{Tr}[A^2+B^2] - \frac{N}{2}
  \text{Tr}[WW^\dagger]} \, 
\ee
with $N=2n+\nu$. The partition function $\tilde Z_{N_f}^\nu$ for the WRMT is then given by 
integrating the matrices $A,B,W$ over the complex Haar measure with
the weight $P(A,B,W)$, i.e.,
\be
{\tilde Z_{N_f}^\nu}(\tilde m, \tilde z; \tilde a_8) = \int  {\cal D}A \, {\cal D}B \,
{\cal D}W \text{det}(\tilde D^\nu_\text{W} + \tilde m + \tilde z \tilde \gamma_5)^{N_f} P(A,B,W) \, . 
\ee

The additional operators introduced by the Wilson lattice discretisation can be
incorporated by extending the partition function using the definition
\be
{\tilde Z_{N_f}^\nu}(\tilde m, \tilde z; \tilde a_6,\tilde a_7,\tilde
a_8) = \int_{-\infty}^\infty dy_6 dy_7 e^{-\frac{y_6^2}{16 |\tilde
    a_6^2|}-\frac{y_7^2}{16 |\tilde a_7^2|}}  {\tilde
  Z_{N_f}^\nu}(\tilde m-y_6, \tilde z-y_7; \tilde a_8) \, .
\ee
Finally, the microscopic limit of WRMT is reached in the limit $N\rightarrow
\infty$ while
\be
\tilde m N, \quad \tilde z N, \quad \tilde a_i N^{1/2} \label{eq:vol_scaling}
\ee
are kept fixed. It is therefore natural to think of $N$ as corresponding to the space-time
volume in the quantum field theory setup.
One can show that the WRMT partition function reproduces the leading terms in the $\epsilon$-expansion of WChiPT
\cite{Akemann:2010em} and the corresponding low-energy constants can
be identified as
\be
\tilde m N \propto m\Sigma V, \quad \tilde z N \propto z\Sigma V, \quad \tilde a_i^2 N \propto a^2\,W_i V.\label{eq:vol_scaling_2}
\ee
Note that the WRMT setup described here can only reproduce WChiPT with a fixed choice for the signs
of the low-energy constants. Using the conventions of \cite{Akemann:2010em}, $\tilde a_8$ will
map into a positive $W_8$, while non-zero values $\tilde a_6$ and $\tilde a_7$ imply negative
values for $W_6$ and $W_7$.
From the corresponding partition functions analytic expressions for
the distributions of the eigenvalues of the QCD
Dirac operator in the $\epsilon$-regime can be obtained both in WChiPT \cite{Akemann:2010em}
and in WRMT \cite{Kieburg:2011uf}.

\section{Lattice setup}
\label{sec:lattice_setup}
Before discussing our results we also need to specify 
the setup for our lattice calculation. 
The massive Wilson-Dirac operator can be written as
\begin{equation}\label{eq:DiracMatrix}
     \DW(m) = \onehalf\gamma_\nu(\nabla_\nu+\nabla_\nu^*)
   - \onehalf\nabla_\nu^*\nabla_\nu + m,
\end{equation}
where $\nabla_\nu,\,\nabla_\nu^*$ denote the covariant forward and
backward lattice derivatives, respectively, $\gamma_\nu$ are the Euclidean Dirac
matrices and $m$ is the bare quark mass.
The Wilson-Dirac operator is $\gamma_5$-hermitian,
\be
\DW^\dagger = \gamma_5 \DW \gamma_5
\ee
and hence
\be
\D5 = \gamma_5 \DW(m)
\ee
is hermitian, $\D5^\dagger = \D5$. The dependence of the eigenvalues $\lambda_k^5(m)$ of
$\D5(m)$ on $m$ defines the eigenvalue flow and from first order
perturbation theory one can infer
\be
\label{eq:chirality_definition}
\frac{d\lambda^5_k}{dm} = \langle k | \gamma_5 | k \rangle \, .
\ee
The eigenvalues of $D_\text{W}$ are denoted by
$\lambda_k^\text{W}$. Since the Wilson Dirac operator is non-normal,
i.e.~$[\DW,\DW^\dagger] \neq 0$,
one needs to distinguish the left and right eigenvectors which define
a bi-orthogonal system. As a consequence the eigensystems of $\DW$ and
$\D5$ are related in a non-trivial way, except for the subset of
$\D5$-eigenvalues which are exactly zero for a given value of $m$. In
that case one has
\be
\gamma_5 (\DW + m) \psi = 0 \quad \Longleftrightarrow \quad \DW \psi = -m \psi \, , \label{eq:equiv_g5_zero}
\ee
i.e., a real mode $\lambda_k^\text{W} \in \R$ of $\DW$ corresponds to
a zero mode of $\D5(m=-\lambda_k^\text{W})$ and the chirality of the
mode is given by the derivative of the Wilson flow,
eq.(\ref{eq:chirality_definition}), at $m=-\lambda_k^\text{W}$.

Here we are specifically interested in the chirality distribution of
the real modes $\lambda_k^\text{W}$ of $D^\nu_\text{W}$ defined as
\be
\rho_\chi(\lambda) =  
\left\langle
\sum_{\lambda_k^W \in \mathbb{R}}
\delta(\lambda_k^W + \lambda)\text{sign}(\langle k|\gamma_5|k\rangle ) 
\right\rangle \, 
\ee
and the distribution of the $k$-th eigenvalue $\lambda_k^5(m)$ of
$\D5$ given by
\be
\rho^k_5(\lambda,m) = \left\langle \delta(\lambda_k^5(m) - \lambda) \right\rangle \, .
\ee
These are the distributions which we compare to the analytic
expressions obtained from WRMT. Up to now only the expression for the
full distribution $\rho_5 = \sum_k \rho^k_5$ is known, so the single
eigenvalue distributions $\rho_5^k$ are obtained from WRMT directly by
numerical simulation.

\subsection{Definition of the topological charge}
A crucial ingredient for the proper treatment of the configurations in
the analysis is their correct assignment to the various topological
sectors.  As discussed in \cite{Akemann:2010em} summing up the signs
of the chiralities of the real modes $\lambda_k^W \in \R$ of $\DW$ for a
given gauge field configuration yields an index of the Wilson Dirac
operator,
\be
\label{eq:charge_definition}
\nu = \sum_{\lambda_k^W \in \mathbb{R}} \text{sign}(\langle
k|\gamma_5|k\rangle )\,.  \ee Note that the sum over the real modes is
actually only a sum over the real modes of $\DW(m)$ in the
neighbourhood of $m$, i.e.~it excludes the doubler real modes.  Since
the real modes are in correspondence with the zero modes of $\D5$,
this index is in agreement with the topological charge as defined
through the Wilson flow method
\cite{Itoh:1987iy,Narayanan:1994gw,Edwards:1998sh,Edwards:1998gk,Edwards:1999ra}.
In practise, the Wilson flow yields all the real modes of $\DW$ up to
a cut-off $m_\text{cut}$, so the practical realisation of
eq.(\ref{eq:charge_definition}) is given by
\be
\nu =  \sum_{\lambda_k^W < m_\text{cut}} \text{sign}(\langle
k|\gamma_5|k\rangle ) \, ,
\label{eq:WF_charge_definition}
\ee
where the restriction takes into account only the real modes up to a
certain cut-off $m_\text{cut}$. Consequently all real modes with
$\lambda_k^W > m_\text{cut}$ are considered to correspond to doubler
modes.  In that sense the topological charge is not well defined at
finite lattice spacing -- as it should be -- but becomes so in the
continuum limit. This ambiguity is not a surprise and is reflected in
any definition of the topological charge at finite lattice spacing. In
particular, also the definition of the topological charge via the
index of the overlap Dirac operator \cite{Neuberger:1997fp},
\be
D_\text{o} = 1 + \gamma_5 \, \text{sign}[ \gamma_5 D_\text{K}(-m)] \, ,
\label{eq:def_overlap_operator}
\ee
is ambiguous, since it depends on the mass shift $-m$ entering the
definition of the hermitian kernel Dirac operator $\gamma_5
D_\text{K}$. The correspondence between the
index of the overlap operator and the charge from the eigenvalue
flow is exact, so $m$ in the overlap operator corresponds to
$m_\text{cut}$ in the definition
eq.(\ref{eq:WF_charge_definition}). 

While the chirality of the real modes is $\pm 1$ in the continuum, for
the non-normal Wilson Dirac operator this holds only approximately
(cf.~also \cite{Kerler:1999dk, Kerler:1999iv}).
The non-normality of $\DW$ is a consequence of its specific symmetry
structure and can not be avoided, however, one can reduce it, e.g., by
modifying the covariant derivatives. In our study we use covariant
derivatives with various levels of HYP-smeared gauge fields, since
this is a simple but efficient mean of suppressing the non-normality
\cite{Durr:2005an}. The effect of the smearing is manifold. Most
importantly, it improves the separation of the physical zero modes
from the doubler zero modes and in this way reduces the ambiguity of
the charge definition, i.e.~its dependence on $m_\text{cut}$.  As a
side effect, the chirality of the zero modes is improved significantly
with increasing smearing level.

Obviously, any remaining ambiguity in the topological charge
assignment is due to additional crossings beyond the value of
$m_\text{cut}$ employed in the charge definition. Interestingly, WRMT
can make predictions about the number of these additional crossings
\cite{Kieburg:2011uf} and we will investigate this in
section~\ref{subsec:additional}.

Based on these considerations, we present in the following the
results for a Wilson Dirac operator with covariant derivatives based
on gauge fields with 5 HYP-smearing levels, and use
eq.(\ref{eq:WF_charge_definition}) as our definition of the
topological charge with $m_\text{cut}$ as given in table
\ref{tab:Nf0_simulation_parameters}.

\subsection{Simulation parameters}
\begin{figure}[!th]
   \centering
   \includegraphics[width=0.7\textwidth]{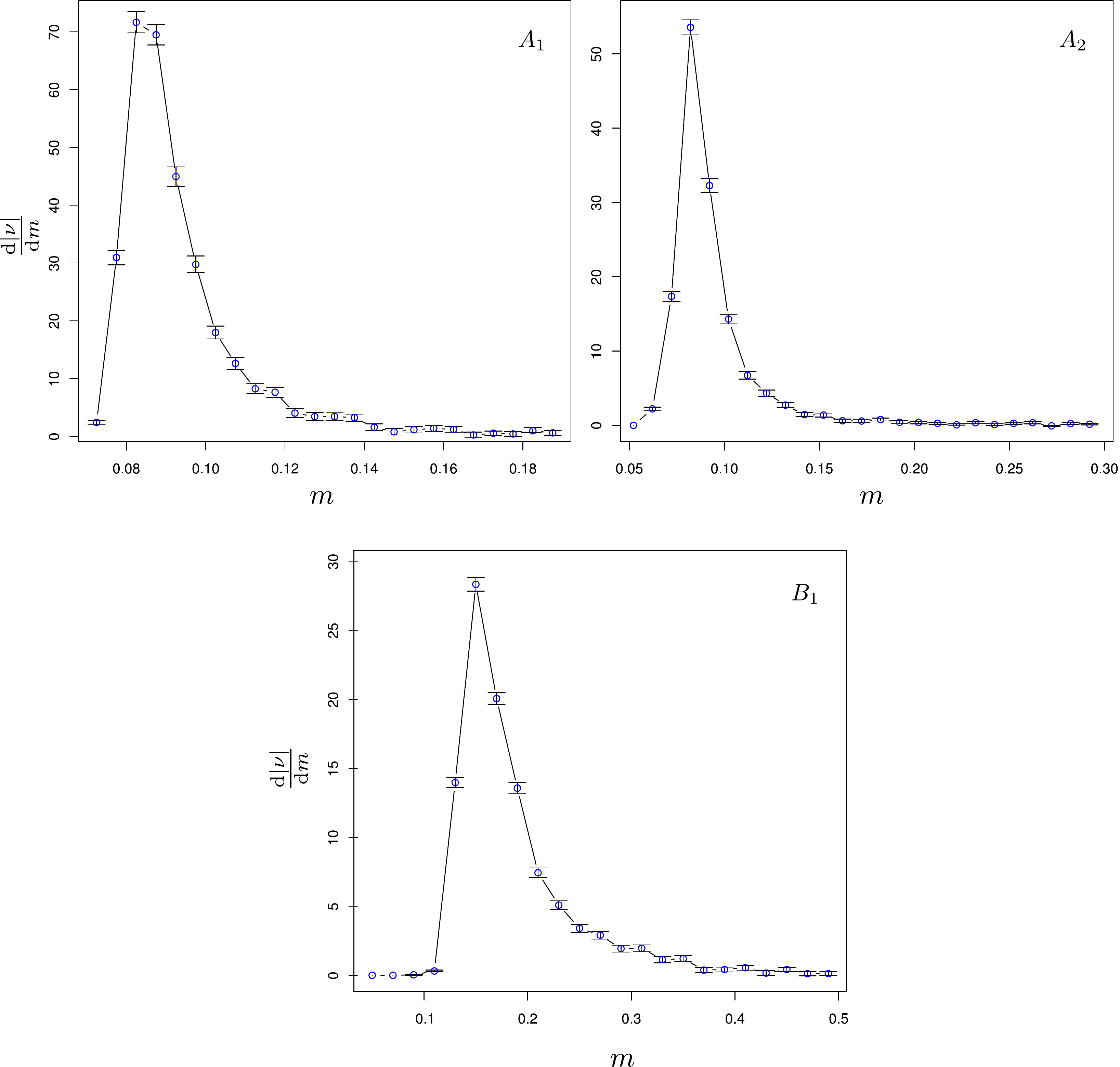}
   \caption{Derivative of the absolute charge with respect to $m_\text{cut}$ for the ensembles described
            in table~\ref{tab:Nf0_simulation_parameters}.}
   \label{fig:charge_derivative_eA}
\end{figure}

We have simulated several lattices with various sizes and lattice
spacings in $N_f=0$ QCD using the Wilson gauge action.  The simulation
parameters for these quenched ensembles are described in table
\ref{tab:Nf0_simulation_parameters}. Within the group $A$ we have kept
the lattice spacing constant in order to study potential finite volume
effects. In contrast, for ensembles $A_1$ and $B_1$ the volume
$V=L^3\times L_t$ is
roughly kept constant in physical units.
\begin{table}[h]
\centering
\begin{tabular}{@{} cccccccc @{}} 
  \toprule
lattice &  $\beta$ & $L/a$ & $L_t/a$ & $r_0/a$ & $L
[\text{fm}]$ &  $m_\text{cut}$ & $\mathrm{d}|\nu|/\mathrm{d}m\big|_{m_\text{cut}}$\\
  \midrule
  $A_1$ &  6.20 & $24$ & $24$ & 7.36 & 1.53 & 0.19 & 0.60(40) \\
  $A_2$ &  6.20 & $20$ & $20$ & 7.36 & 1.28 & 0.30 & 0.14(11) \\
  $B_1$ &  5.90 & $14$ & $16$ & 4.48 & 1.47 & 0.50 & 0.12(13) \\
\bottomrule
\end{tabular}
\caption{Simulation parameters for the $N_f=0$ ensembles together with
  the value of $m_\text{cut}$ used in the definition of the
  topological charge.}
\label{tab:Nf0_simulation_parameters}
\end{table} 
Table~\ref{tab:Nf0_simulation_parameters} also provides a numerical
estimate of the derivative of the charge with respect to the cut
employed in the Wilson flow. A graphical presentation of this quantity
as a function of the mass is shown for each of the ensembles in figure~\ref{fig:charge_derivative_eA}. 
The derivative in each of the ensembles shows a sharp peak in the
neighbourhood of the critical mass, after which a more or less rapid
decay sets in depending on the volume and the lattice spacing. In
absolute terms, the derivative of the charge is practically negligible
at the point where the mass cut is introduced. Using a
linear extrapolation through the tail of the distributions as a rough
estimate, we should expect to be misassigning up to half a percent of
the configurations in ensemble $B_1$ and at most about one percent for
the ensembles $A_1$ and $A_2$.

\section{Results}

\subsection{Cumulative hermitian eigenvalue distributions}
\begin{figure}[!h]
   \centering
   \includegraphics[width=\textwidth]{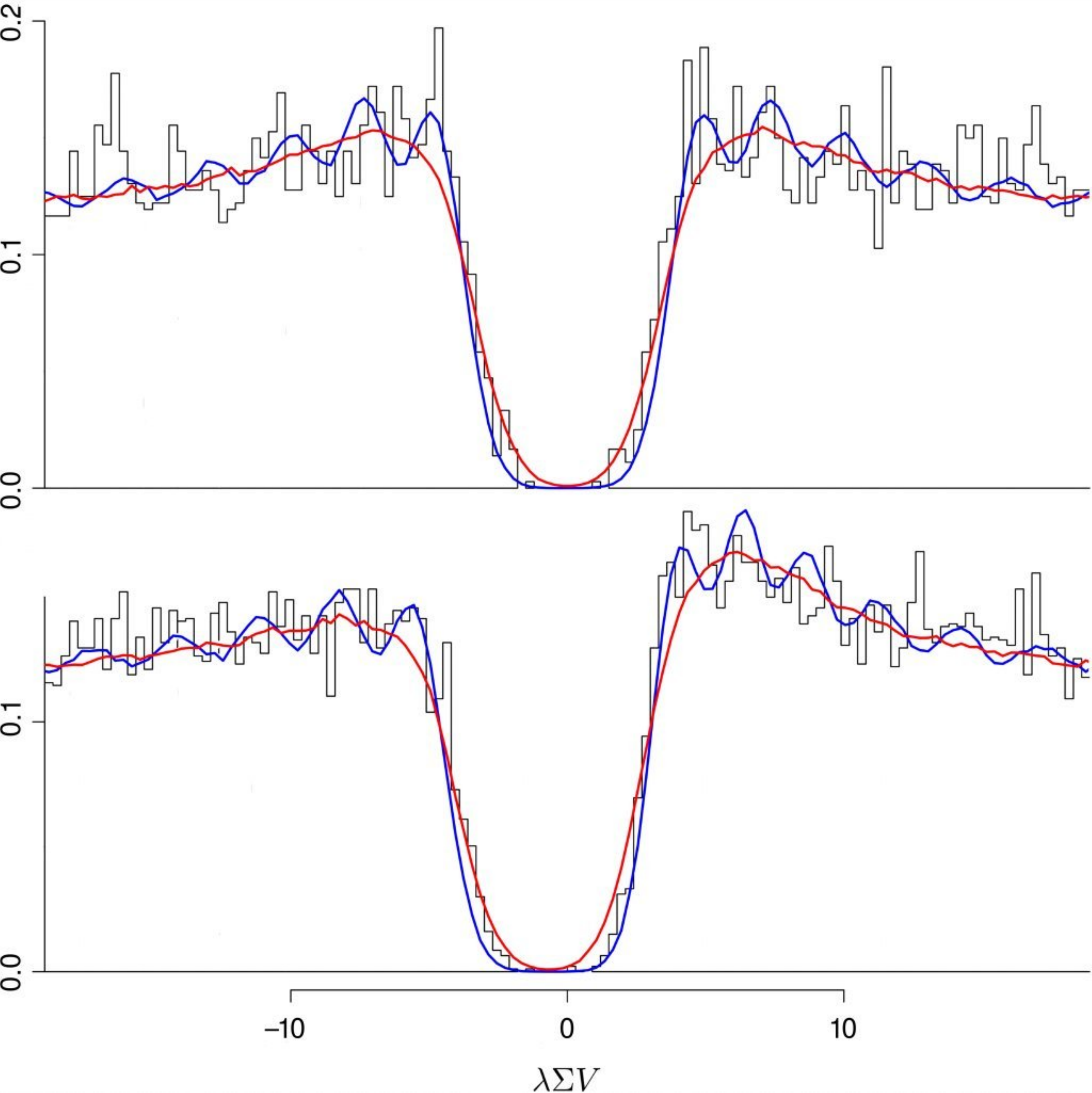}
   \caption{The density of the lowest lying eigenvalues in the
     topological charge sectors $\nu=0$ (top) and $\nu=1$ (bottom), for the
     ensembles $A_1$. Blue lines are fits accounting just for $W_8$, red lines
     are fits including the effects of $W_6$ and $W_7$. These lines were calculated
     by means of WRMT, using 250k samples.}
   \label{fig:spectrum_overview}
\end{figure}

We start by considering the distributions $\rho_5(\lambda)$ of the
eigenvalues of the hermitian Wilson Dirac operator $\D5$ and compare
them with the analytic expressions. So far the expressions from WRMT
and WChiPT comprise only the full distributions for the lowest lying
eigenvalues. In figure~\ref{fig:spectrum_overview} we show the full
distributions for our ensemble $A_1$ in the charge sectors $\nu=0$ and $\nu=1$,
together with fits to both that will be discussed in more detail below.  
Some of the expected features of the spectrum are apparent, the most prominent
being the enhanced peak on the positive side for $\nu=1$. However,
it is clear
 -- also from a comparison of the data to the relatively delicate
features of the fitted lines -- 
that it would be rather challenging to push the analysis beyond a
qualitative comparison on the basis of these distributions alone.
Luckily, our data obtained from the simulations contain much more
information than just the full distribution, since we have access to
each single low-lying eigenvalue.  Indeed it turns out that the shapes
of the single eigenvalue distributions are rather sensitive to the
low-energy constants $W_6, W_7$ and $W_8$. For this reason, our fits
concern single eigenvalue distributions obtained from explicit WRMT
simulations including the effects of $W_8$ and possibly $W_6,
W_7$. The fit parameters in WRMT are then $\tilde m, \tilde a_6,
\tilde a_7, \tilde a_8$ and $\Sigma V$ which sets the scale and
connects the WRMT to the QCD data.

Binning the data into histograms introduces an unwelcome dependence on
the bin size and hence we choose to compare cumulative
distributions. In order to assess the quality of the fits we employ
the Kolmogorov-Smirnov (KS) test statistics. If $S_\text{QCD}(x)$ and
$S_\text{WRMT}(x)$ are the estimates of the cumulative distribution
functions obtained from the QCD and the WRMT simulation, respectively,
the KS statistics is given by
\be
\label{eq:def_KS_statistics}
d_\text{max} = \operatorname*{max}_{-\infty < x < \infty} |S_\text{QCD}(x) - S_\text{WRMT}(x)| \, .
\ee
Since we fit several distributions at once, it is not clear how to
translate $d_\text{max}$ into a quality of fit measure and we will
restrict ourselves to reporting only the value of $d_\text{max}$.

Finally, the errors on the distributions we present in the following
are obtained by a bootstrap procedure and for better visibility
represent a 95\% confidence bound, rather than a regular standard
deviation.

A summary of all the fit results discussed in this
paper is given in table~\ref{tab:fit_results} for convenient
reference.
\begin{table}[h]
\centering
\begin{tabular}{@{} cccrcccccccc @{}} 
  \toprule
& ens. &  $|\nu|$ & $N_\text{conf}$ & $m$ & $N
\tilde m$ & $\tilde a_6 N^{1/2}$ & $\tilde a_7 N^{1/2}$ & $\tilde a_8
N^{1/2}$ & $\Sigma V$ & $d_\text{max}$ \\
  \midrule
$F_1$   & $A_1$ & 1 & 1602 &  -0.07 & 5.3(2) & 0.25(7) & 0.25(5) &  0.70(13)  & 308.3(5.8) &   0.061   \\
&&&&&&&&&&&\\                                                                                              
$F_2$   & $A_1$ & 0 &  462 &  -0.07 & 5.5(4) & 0.27(4) & 0.27(2)  &  0.82(13)  & 299.6(3.4) &   0.064 \\
$F_3$   & $A_1$ & 1 & 1602 &  -0.07 & 5.5(2) & 0.23(9) & 0.27(2)  &  0.83(06)  & 300.4(2.2) &   0.087 \\
$F_4$   & $A_1$ & 2 &  612 &  -0.07 & 5.6(3) & 0.18(7) & 0.27(5)  &  0.84(13)  & 301.1(8.6) &   0.152 \\
&&&&&&&&&&&\\                                                                                              
$F_5$   & $A_1$ & 0 &  462 &  -0.07 & 5.4(--) & 0        & 0         &  0.85(--)  & 295.2(--)  &   0.091 \\
$F_6$   & $A_1$ & 1 & 1602 &  -0.07 & 5.7(--) & 0        & 0         &  0.96(--)  & 303.7(--)  &   0.155 \\
&&&&&&&&&&&\\
$F_7$   & $A_1$ & 0 &  462 &  -0.055& 8.8(8)   & 0.22(7) & 0.22(4)  & 0.63(19)  & 296.3(21.9)&   0.149   \\
$F_8$   & $A_1$ & 1 & 1602 &  -0.055& 8.2(8)   & 0.23(5) & 0.23(9)  & 0.51(19)  & 282.8(9.8) &   0.129   \\
&&&&&&&&&&&\\                                                                                                                                           
$F_9$   & $A_2$ & 0 &  601 &  -0.07 & 4.0(1.6) & 0.20(4) & 0.18(4)  & 0.76(32)  & 156.9(21.5) &   0.196   \\
$F_{10}$ & $A_2$ & 1 & 1558 &  -0.07 & 4.9(2.5) & 0.32(5) & 0.29(5)  & 0.77(26)  & 170.8(21.4) &   0.346   \\
&&&&&&&&&&&\\
$F_{11}$ & $B_1$ & 0 &  624 &  -0.1  & 7.6(7)   & 0.29(20) & 0.29(13)  & 0.82(13)  & 150.9(5.7) &    0.091  \\
$F_{12}$ & $B_1$ & 1 & 1241 &  -0.1  & 6.8(7)   & 0.29(14) & 0.27(14)  & 0.77(32)  & 141.4(4.9) &    0.260  \\
\bottomrule
\end{tabular}
\caption{Fit results for the fits of WRMT to the QCD ensembles $A_1,
  A_2$ and $B_1$ in the various topological sectors $|\nu|$. $N_\text{conf}$ denotes the number of
  configurations used in the fits. The bare mass $m$ of the Dirac
  operator and $\Sigma V$ are given in lattice units, while the
  remaining columns are the fitted dimensionless
  WRMT parameters. The total number of
  the lowest-lying eigenvalues fitted on both the positive and
  negative side of the spectrum is $N_\text{ev}=4$ for fit $F_1$ and
  $N_\text{ev}=8$ for all other fits.}
\label{tab:fit_results}
\end{table}

\subsubsection{Influence of $W_6$ and $W_7$}

\begin{figure}[!ht]
   \centering
   \includegraphics[width=\textwidth]{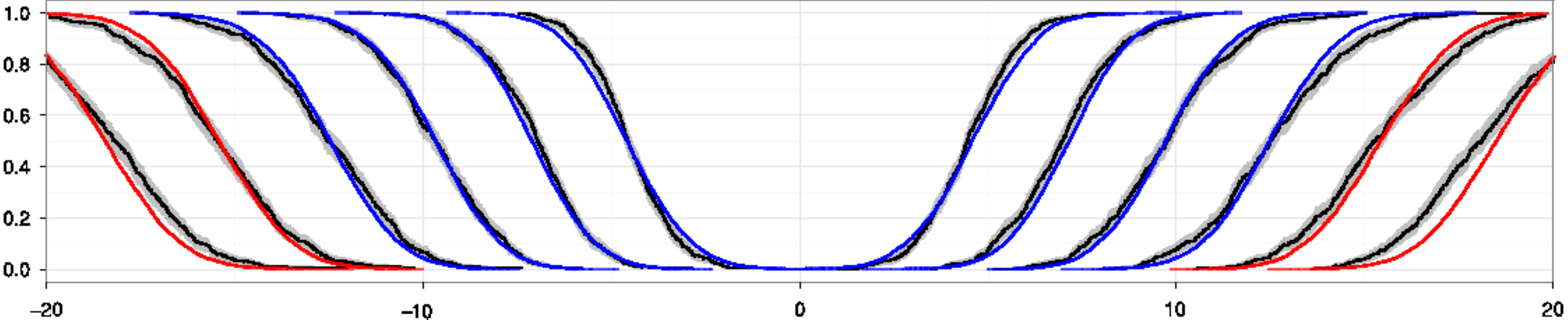}\\
   \includegraphics[width=\textwidth]{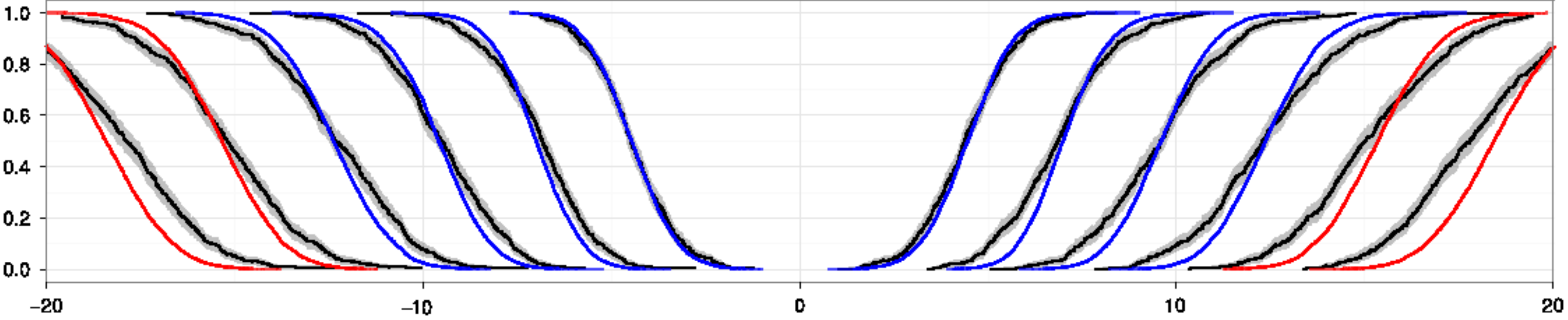}\\
   \centering{$\lambda \Sigma V$}
   \caption{Cumulative eigenvalue distributions for ensemble $A_1$ in
     charge sector $\nu=0$ compared to WRMT fits including
     contributions from low-energy constants $W_6, W_7, W_8$ (fit $F_2$,
     upper plot) and only $W_8$ (fit $F_5$, lower plot). Blue lines denote
     fitted distributions from WRMT, red lines are the predictions
     following from the fit.}
   \label{fig:W678_vs_W8}
\end{figure}

As mentioned above, large-$N_c$ arguments indicate that the terms
proportional to $W_6$ and $W_7$ are suppressed~\cite{Kaiser:2000gs}.
In order to assess the importance of these low-energy constants we
compare fits including only $W_8$ (fits $F_5$ and $F_6$ in table
\ref{tab:fit_results}) to ones which include all three low-energy
parameters $W_6, W_7$ and $W_8$ (fits $F_2$ and $F_3$). The resulting
distributions obtained in ensemble $A_1$ in charge sector $\nu=0$ at
$m=-0.07$ are displayed in figure \ref{fig:W678_vs_W8} and a visual
examination makes it clear that the fits including $W_6, W_7$ are
superior in describing the shapes of the single eigenvalue
distributions. This is corroborated by the values of $d_\text{max}$ in
table \ref{tab:fit_results} which grow by about 30-50\% when $W_6,
W_7$ are excluded. So while $W_8$ appears to account for large parts
of the deviation from the chiral setup, we do find $W_6$ and $W_7$ to
be important as well and henceforth include them in all further fits.

Fits including and excluding the effects of $W_6$ and $W_7$ have also 
been added to figure~\ref{fig:spectrum_overview}. Here one can clearly
observe the influence of the additional operators, that work to largely
wash out the characteristic but subtle undulating pattern observed on top of 
the spectrum as predicted by the WChiPT expressions~\cite{Akemann:2010em}.
Our observation that fitting these added parameters tends to produce
non-vanishing values implies that one should not necessarily expect
to observe such a pattern in practise.
This once again underlines the importance of examining the distributions of the
separate eigenvalues rather than the overall distribution, as the apparent
impact of these operators would imply few distinct features of the spectrum 
remain to constrain fits by.

\subsubsection{Sensitivity to topology and predictive power}

Next we investigate the sensitivity of the WRMT fits to the various
topological sectors and the predictive power of the fits as a first
non-trivial test. A possible way to proceed is to fit the
distributions of a selected set of eigenvalues in a given topological
sector to the distributions obtained from WRMT and check the
consistency for the other eigenvalues and in the other topological
sectors.

\begin{figure}[!ht]
   \centering
\includegraphics[width=\textwidth]{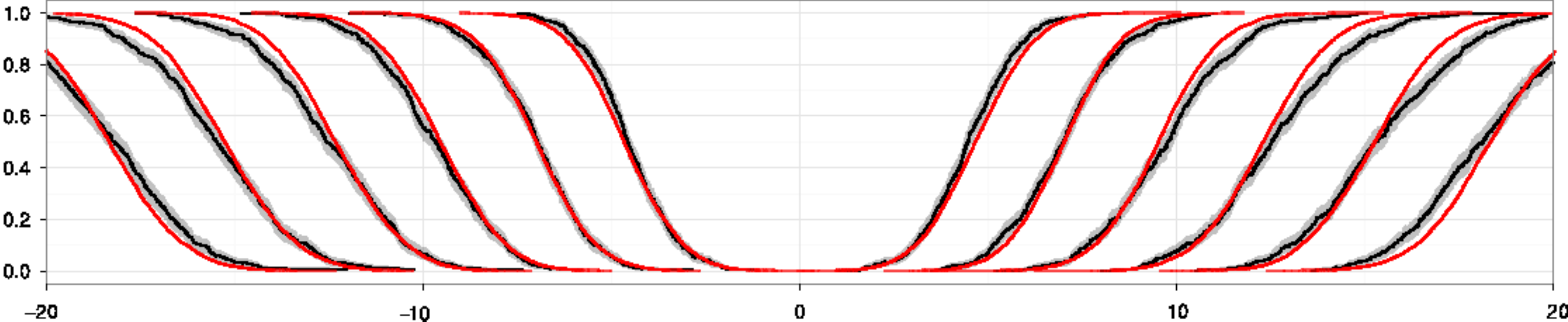}\\
\includegraphics[width=\textwidth]{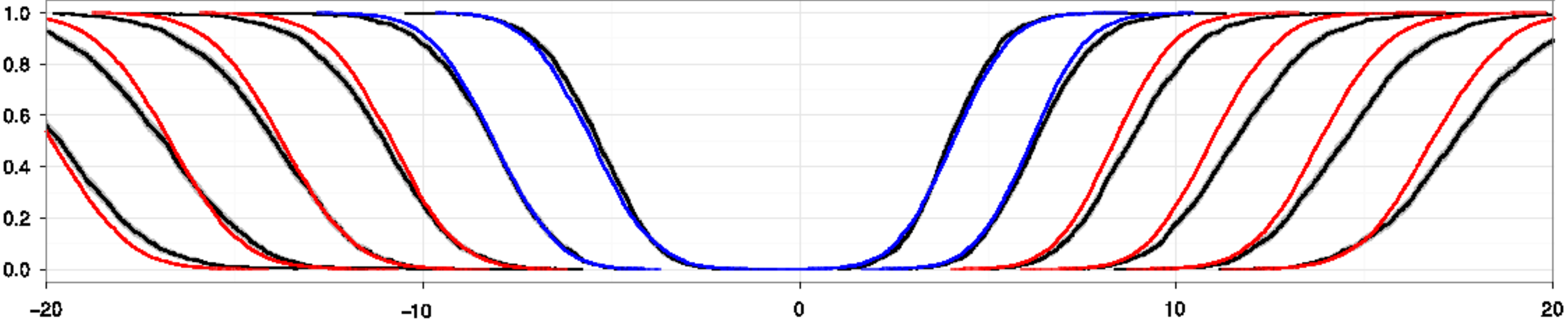}\\
\includegraphics[width=\textwidth]{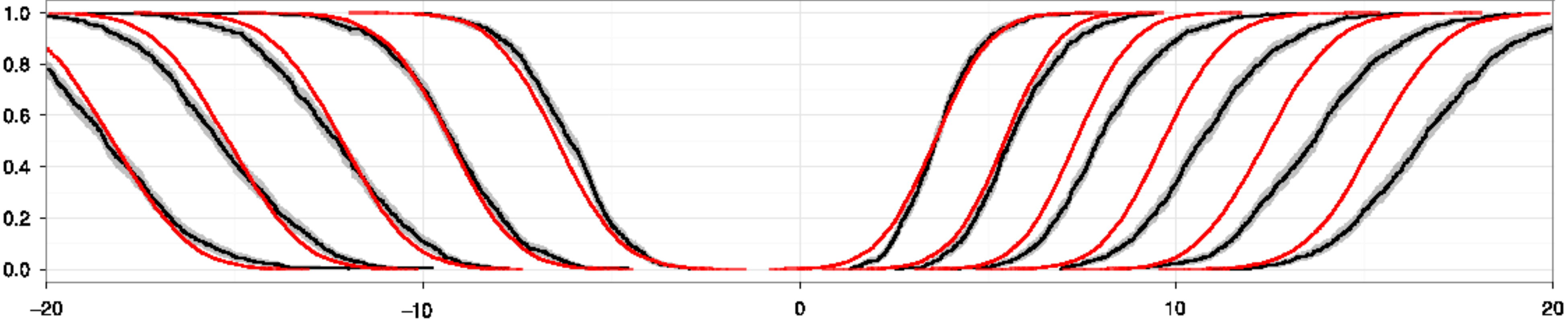}\\
   \centering{$\lambda \Sigma V$}
   \caption{Cumulative eigenvalue distributions for ensemble $A_1$ in
     charge sectors $\nu=0, 1$ and 2 from top to bottom. The QCD data
     is displayed in black. Blue lines denote the fitted distributions
     from WRMT (fit $F_1$ in table~\ref{tab:fit_results}), while red
     lines are the predictions following from the fit. }
   \label{fig:single_ev_fit_A1}
\end{figure}

In figure \ref{fig:single_ev_fit_A1}
we show the results of such an exercise for the spectrum of $\D5$ at a
bare mass $m=-0.07$. We choose to fit the distributions of the lowest
two positive and negative eigenvalues in the charge sector $\nu=1$,
since this is where we have the largest statistics. As before, the
fitted cumulative WRMT distributions are displayed as the blue lines,
while the red lines are the predictions for the remaining eigenvalue
distributions following from the fit. The results for the fit
parameters and the value of $d_\text{max}$ can be found in
table~\ref{tab:fit_results} in the row labelled by $F_1$.

A number of observations can be made at this point:

\noindent(1) The four eigenvalue distributions in the charge sector
$\nu=1$ can be described very well by WRMT, both in terms of the
position and the shape, as reflected by the low value of
$d_\text{max}$ reported in table~\ref{tab:fit_results}.\\
\noindent(2) The distributions in the charge $\nu=0$ sector are all
predictions. Again, the agreement is very good up to about $|\lambda|
\Sigma V \simeq 12$ where deviations in the tail of the distributions
away from 0 start to show up. One may assign this to the onset of the bulk
which could be related to the Thouless energy where the RMT
description is expected to break down.\\ 
\noindent(3) For the predicted distributions in the charge $\nu=1$
sector we find that WRMT tends to underestimate the tails of the
eigenvalue distributions away from 0.\\
\noindent(4) The same effect is visible in the other predicted
eigenvalue distributions in charge sector $\nu=2$, but in addition we
now also find a mismatch in the position of the
distributions.\\
\noindent(5) It seems that both these mismatches are more severe on
that side of the spectrum where the would-be real modes are
accumulating. One may conclude from this that lattice artefacts seem
to be pushed from the would-be real modes to the nearby modes in the
spectrum, and that this effect is enhanced for increasing topology.

\subsubsection{Scaling of the distributions}

An important prediction of the WRMT is the universal scaling of the
distributions with the volume, as implied by equation
\ref{eq:vol_scaling_2}, and with the mass.

\begin{figure}[!t]
   \centering
   \includegraphics[width=\textwidth]{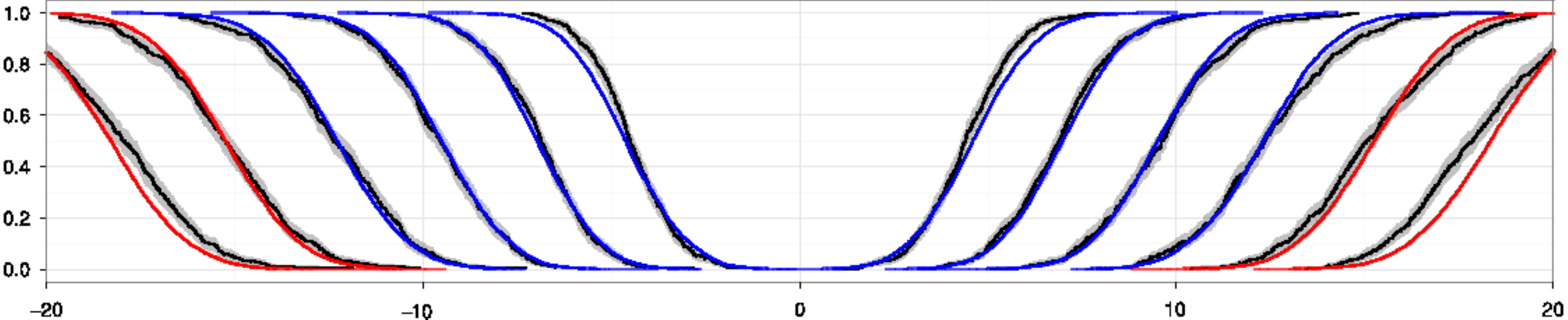}\\
   \includegraphics[width=\textwidth]{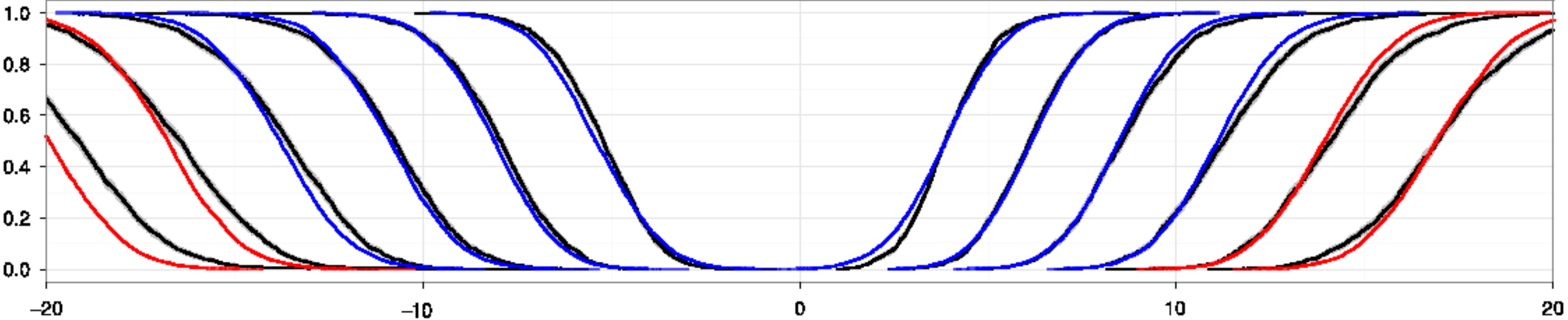}\\
   \centering{$\lambda \Sigma V$}
  \caption{Cumulative eigenvalue distributions for ensemble $A_1$ with
    volume $V_1=(1.52\text{fm})^4$ together with the best fits to
    WRMT. Charge sector $\nu=0$ (top, fit $F_2$) and $\nu=1$ 
    (bottom, fit $F_3$).}
   \label{fig:basic_fits}
\end{figure}

\begin{figure}[!b]
   \centering
   \includegraphics[width=\textwidth]{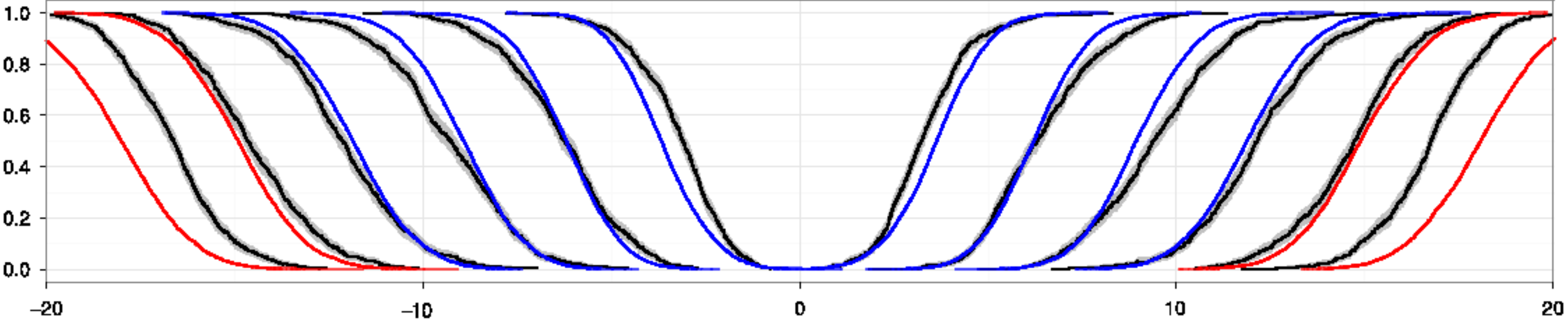}\\
   \includegraphics[width=\textwidth]{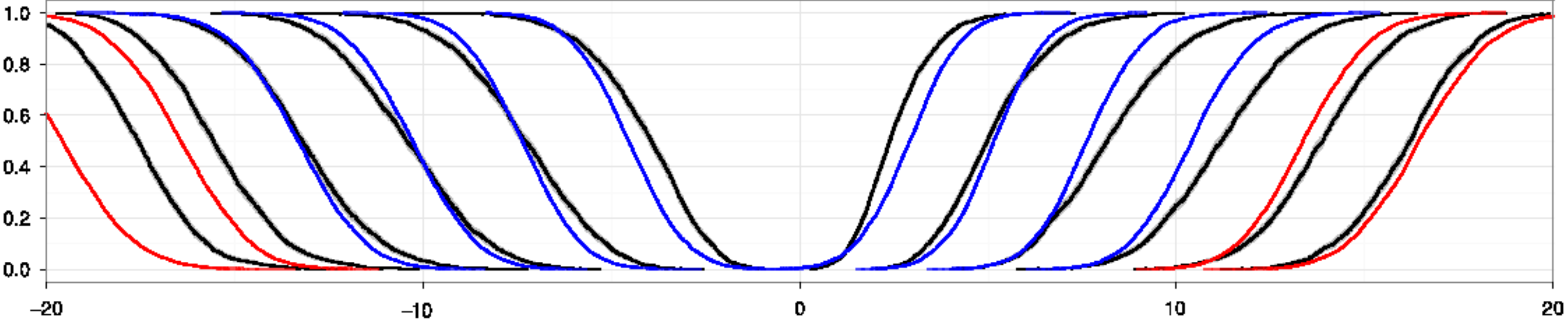}\\
   \centering{$\lambda \Sigma V$}
  \caption{Cumulative eigenvalue distributions for ensemble $A_2$  
    with $V_2=(1.28\text{fm})^4$ together with the best fits to WRMT. 
    Charge sector $\nu=0$ (top, fit $F_9$) and $\nu=1$ (bottom, fit $F_{10}$).}
   \label{fig:volume_comparison}
\end{figure}

We first investigate the scaling of the distributions with the
volume. For this purpose we compare our best fits $F_{2-3}$ on
ensemble $A_1$ with a volume $V_1=(1.52\text{fm})^4$ in the charge
sectors $\nu=0, 1$ to the corresponding fits $F_{9-10}$ on ensemble
$A_2$ with volume $V_2=(1.28\text{fm})^4$. Both data sets are obtained
at $m=-0.07$. The fits are displayed for visual inspection in
figure~\ref{fig:basic_fits} for $A_1$ and
figure~\ref{fig:volume_comparison} for $A_2$. We note that
the quality of the fits on the smaller volume is clearly
deteriorating, with $d_\text{max}$-values increasing by a factor 3 in
charge sector $\nu=0$ and 4 in sector $\nu=1$.  We assign this to the
fact that the volume $V_2=(1.28\text{fm})^4$ is clearly too small for
the system to be in the $\epsilon$-regime. Nevertheless, with the
results in table \ref{tab:fit_results} we may quantitatively check the
scaling of the parameters. We expect
$\tilde a_6, \tilde a_7$ and $\tilde a_8$ to scale like $(V_1/V_2)^{1/2}=1.41$
and $\Sigma V$ like $V_1/V_2=2.0$. Except for $\tilde a_8$ this is what we
roughly observe at least in the charge sector $\nu=0$ where the small
volume is indeed better described by WRMT.

\begin{figure}[!ht]
   \centering
   \includegraphics[width=\textwidth]{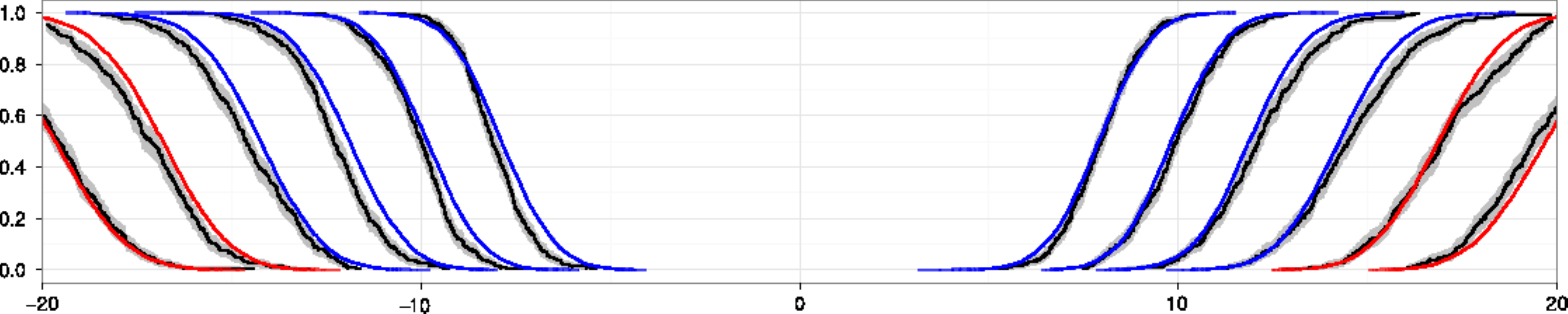}\\
   \includegraphics[width=\textwidth]{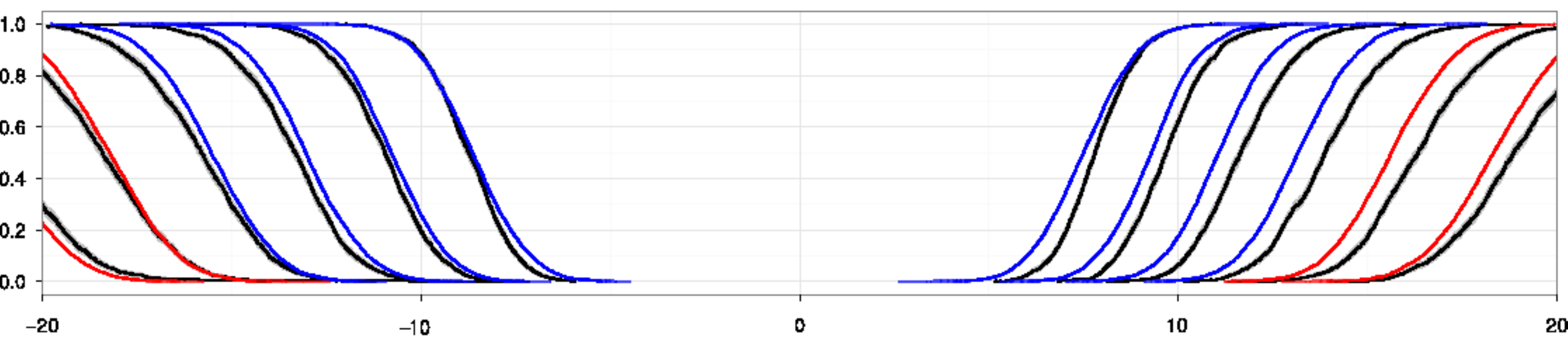}\\
   \centering{$\lambda \Sigma V$}
  \caption{Cumulative eigenvalue distributions for ensemble $A_1$ at
    bare mass $m=-0.055$ in charge sectors $\nu=0$ (top, fit $F_7$) and $\nu=1$ 
   (bottom, fit $F_8$).} \label{fig:m055_fits}
\end{figure}

Next we investigate the scaling of the distributions with the mass. For
this purpose we plot in figure \ref{fig:m055_fits} the
distributions obtained on $A_1$ in the charge sectors $\nu=0, 1$ at a different
value of the bare mass of $m=-0.055$. These fits should again be compared
with those in figure~\ref{fig:basic_fits}.
From a comparison of the fit parameter $\tilde m$, we can infer a
value of the critical mass $m_\text{crit} \sim -0.09$. As we will see in section
\ref{sec:chirality_distributions} this matches rather well with what
we find from the distributions of the real modes.  On the other hand,
we observe a rather poor quality of the fits $F_7$ and $F_8$ as
compared to $F_2$ and $F_3$.  Of course, the increase in mass will
push all the eigenvalues in the spectrum towards the bulk making it
more difficult for the WRMT to describe them properly. Indeed, this is
also reflected in the enhanced errors on all the fit parameters
impeding a more quantitative comparison of the two spectra at
different masses.

\subsection{Eigenvalue averages}

\begin{figure}[!ht]
   \centering
   \includegraphics[width=\textwidth]{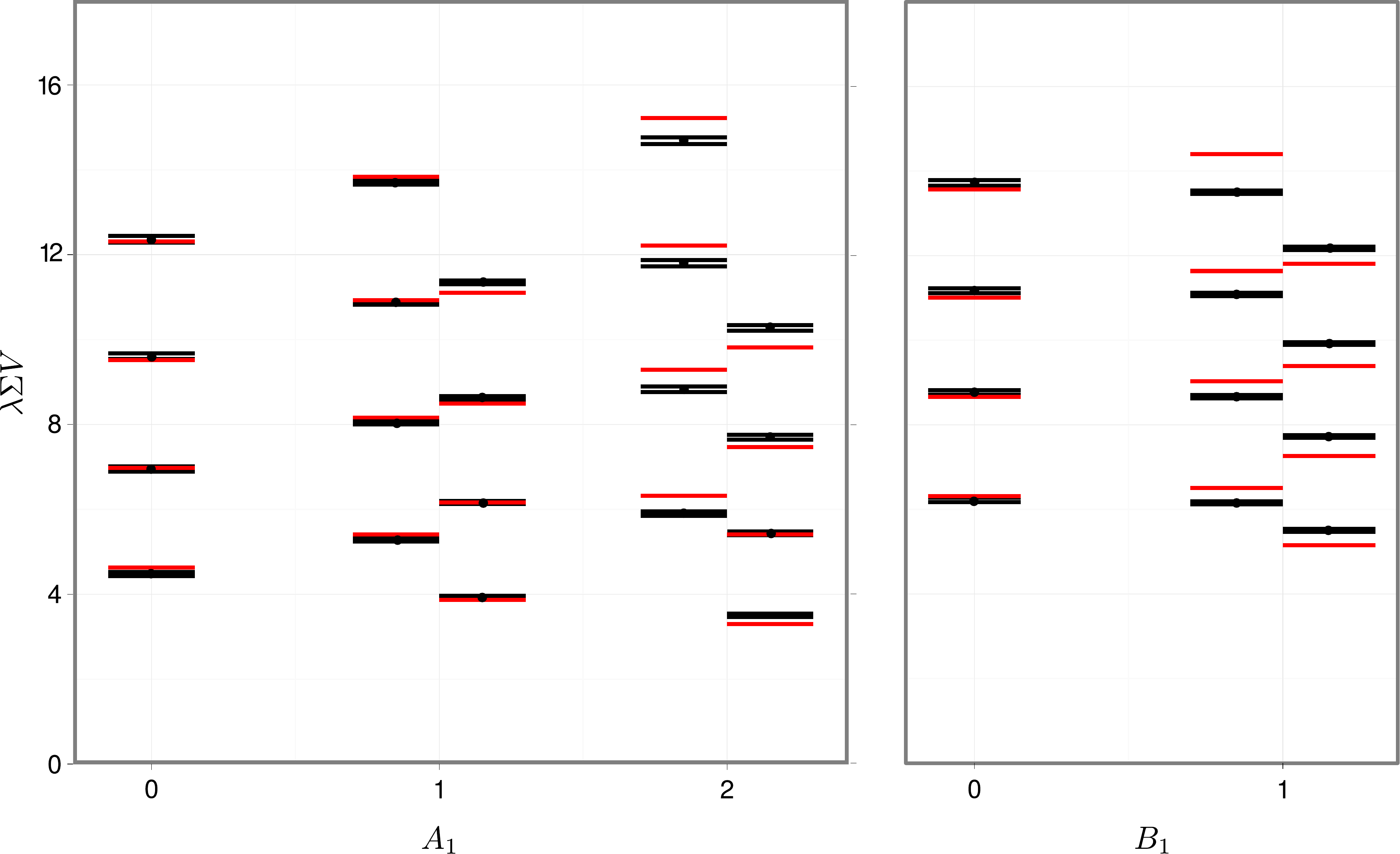}
   \caption{Comparison of the measured and fitted eigenvalue averages,
     for the ensembles $A_1$ and $B_1$ in the different topological
     charge sectors. Negative eigenvalues are plotted to the left,
     positive eigenvalues to the right. Black points indicate the
     measured data, while red lines indicate the values from
     WRMT based on the fits $F_2$, $F_3$, $F_4$,
     $F_{11}$ and $F_{12}$ from table~\ref{tab:fit_results}.}
   \label{fig:sb62_24_24_nu}
\end{figure}

With respect to the full eigenvalue distributions discussed in the
previous section, averages of single eigenvalues represent a strong
reduction of the amount of information available for a
comparison. Nevertheless, since the averages can be determined to high
statistical precision, a comparison between the QCD data and the WRMT
description at this level may indicate systematic deviations more
clearly.  Figure~\ref{fig:sb62_24_24_nu} summarises the results of our
fits to the two main ensembles $A_1$ and $B_1$ in terms of these
averages.  The plots show the averages for each separate eigenvalue,
both as measured on the lattice and as produced by the fitted WRMT. In
each topological charge sector, values on the left denote the negative
eigenvalues, while the values on the right give the positive ones. For
the charge sector $\nu=0$ the two sides are symmetric and have been
merged.

We find that the WRMT appears to work very well for trivial topology,
both for the coarse ensemble $B_1$ and the smooth one $A_1$, as can
also be inferred from table~\ref{tab:fit_results}.  For higher
charges, however, the eigenvalues with chiralities opposite to the
sign of the topological charge are systematically pulled towards zero,
while the ones with the same chirality sign come out higher than what
WRMT tends to predict. This points to the mixing between eigenvalue
sectors of different chirality being stronger, and hence lattice
artefacts being larger than what WRMT is able to describe. The effect
seems to be more pronounced in the higher charge sectors and at
coarser lattice spacing. This is corroborated by the fact that for the
ensemble $A_1$ the charge sector $\nu = 1$ is still well described,
whereas for $B_1$ large deviations start to appear already there.

\subsection{Chirality distributions}
\label{sec:chirality_distributions}

The determination of the topological charge through the Wilson flow
provides us with all the zeros of $\D5(-m)$ up to the cut-off
$m_\text{cut}$ together with their associated chirality.  According to
the relationship in eq.(\ref{eq:equiv_g5_zero}), the locations of
these zeros can be directly related to the real eigenvalues
$\lambda_k^\text{W}$ of the operator $\DW$. The latter can also be
directly obtained from WRMT, once the parameters have been determined,
allowing for the comparison of the chirality distributions.

\begin{figure}[!th]
   \centering
   \includegraphics[width=0.7\textwidth]{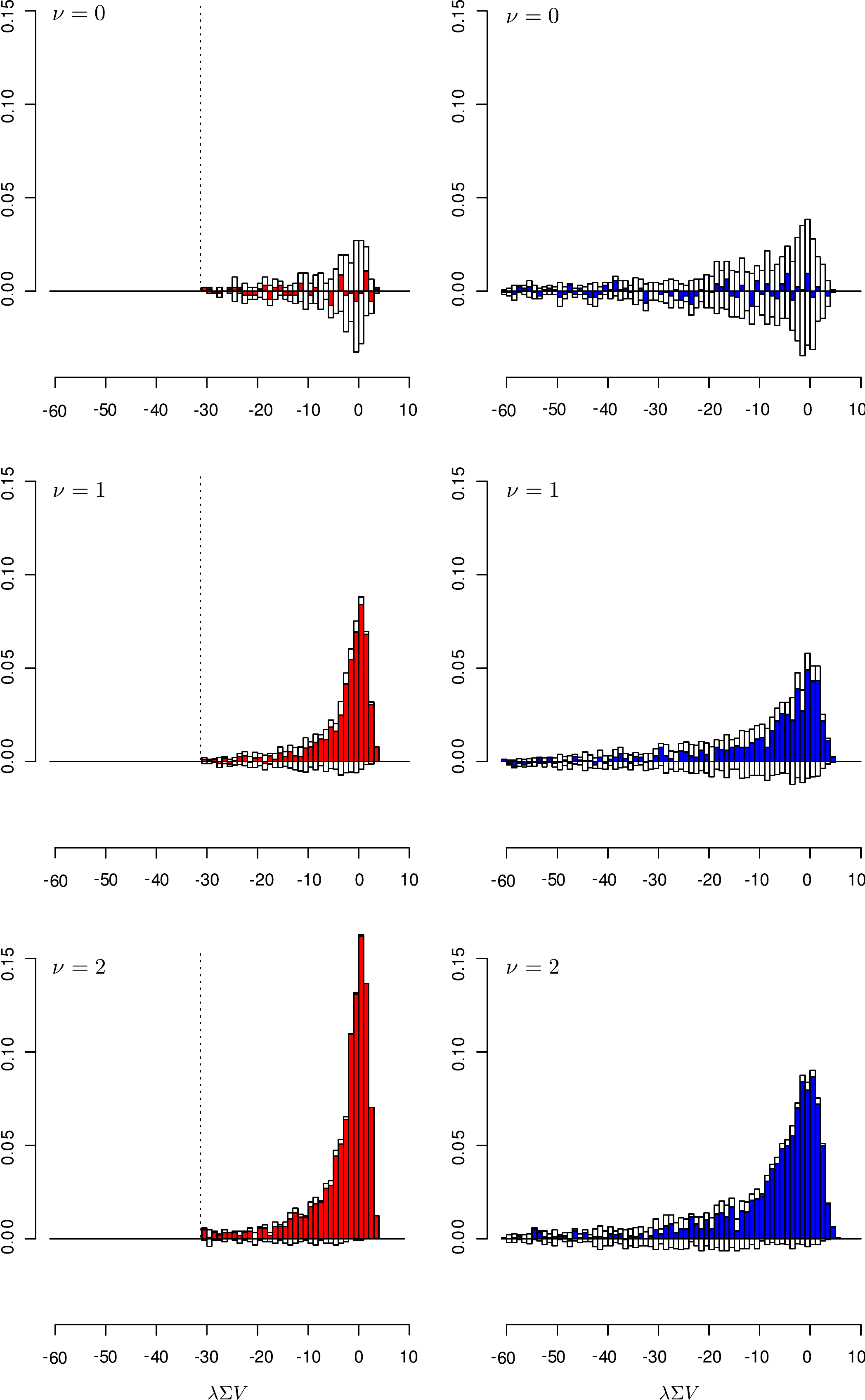}
   \caption{Chirality distributions for the three lowest charge
     sectors of ensembles $A_1$ (red) and $B_1$ (blue).  The
     distributions have been shifted for convenient comparison and
     have been normalised to contain the same statistics per charge
     sector. The dotted line corresponds to the value of
     $m_\text{cut}$ employed for the calculation of the charge in
     ensemble $A_1$.}
   \label{fig:charge_hists}
\end{figure}

Figure~\ref{fig:charge_hists} shows a comparison of the chirality
distributions that were measured for our ensembles $A_1$ (in red) and
$B_1$ (in blue). Empty bars show the number of eigenvalues with
positive and negative chirality, while the filled bars give the net
chirality distribution.  To simplify the comparison, the measured
values of $\lambda \equiv -\lambda_k^\text{W}$ were multiplied by
$\Sigma V$ as determined from our fits and shifted such that the peak
of the distribution is located at 0 in a rough approximation of the
critical mass. The distributions for $B_1$ were rescaled to compensate
for differences in the statistics per charge sector.

The distributions for both ensembles $A_1$ and $B_1$ are sharply
peaked around $m_c$, but long tails remain towards more negative
values of $\lambda$.  It is clear that the distribution is
more sharply peaked, and the tail shorter, for the ensemble $A_1$ with
its significantly smaller lattice spacing.  In addition, the number of
cancellations is significantly higher for the ensemble $B_1$. The
occurrence of conjugate pairs of real modes with opposite chirality,
responsible for these cancellations, will be suppressed towards the
continuum limit.  This comparison therefore shows a clear decrease in
discretisation artifacts.

\begin{figure}[!th]
   \centering
   \includegraphics[width=0.5\textwidth]{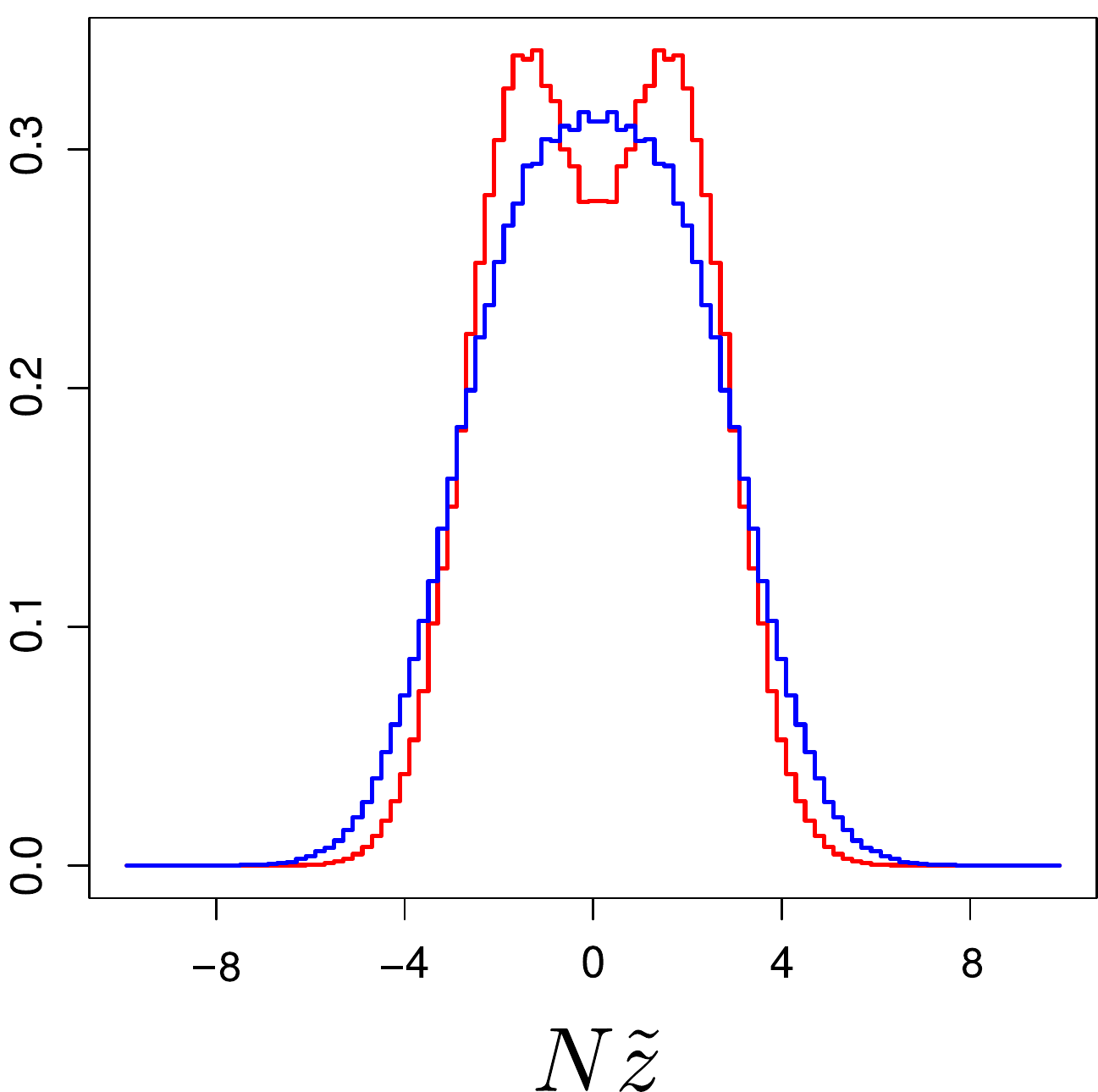}
   \caption{Chirality density as calculated by WRMT for $\nu=2$,
     excluding (red) and including (blue) the effect of $\tilde a_6$
     and $\tilde a_7$.}
   \label{fig:densities_w67_effect}
\end{figure}

Analytical results for the chiral density~\cite{Akemann:2010em} based
on WChiPT tend to show a distinct structure of merged peaks,
the number of which is given by the topological charge. This is in contrast 
to what we observe in any of our ensembles. One possible cause of this could 
be the influence of the $W_6$ and $W_7$ operators, as already indicated in 
figure~\ref{fig:spectrum_overview}. To investigate this further, we obtained
the real eigenvalues from 100k samples of the WRMT with $\nu=2$. We
used sets of parameters matching our $A_1$ ensemble, both
including $\tilde a_6$ and $\tilde a_7$. The chirality distributions
obtained from this calculation are provided in
figure~\ref{fig:densities_w67_effect}.
The effect of the additional operators is clearly observable, as the
peak is broadened and all internal structure wiped out. Given the
non-vanishing values we find for the parameters $\tilde a_6$ and
$\tilde a_7$ in our fits, we should not expect to see any structure in
our chirality distributions.

\begin{figure}[!t]
   \centering
   \includegraphics[width=7cm]{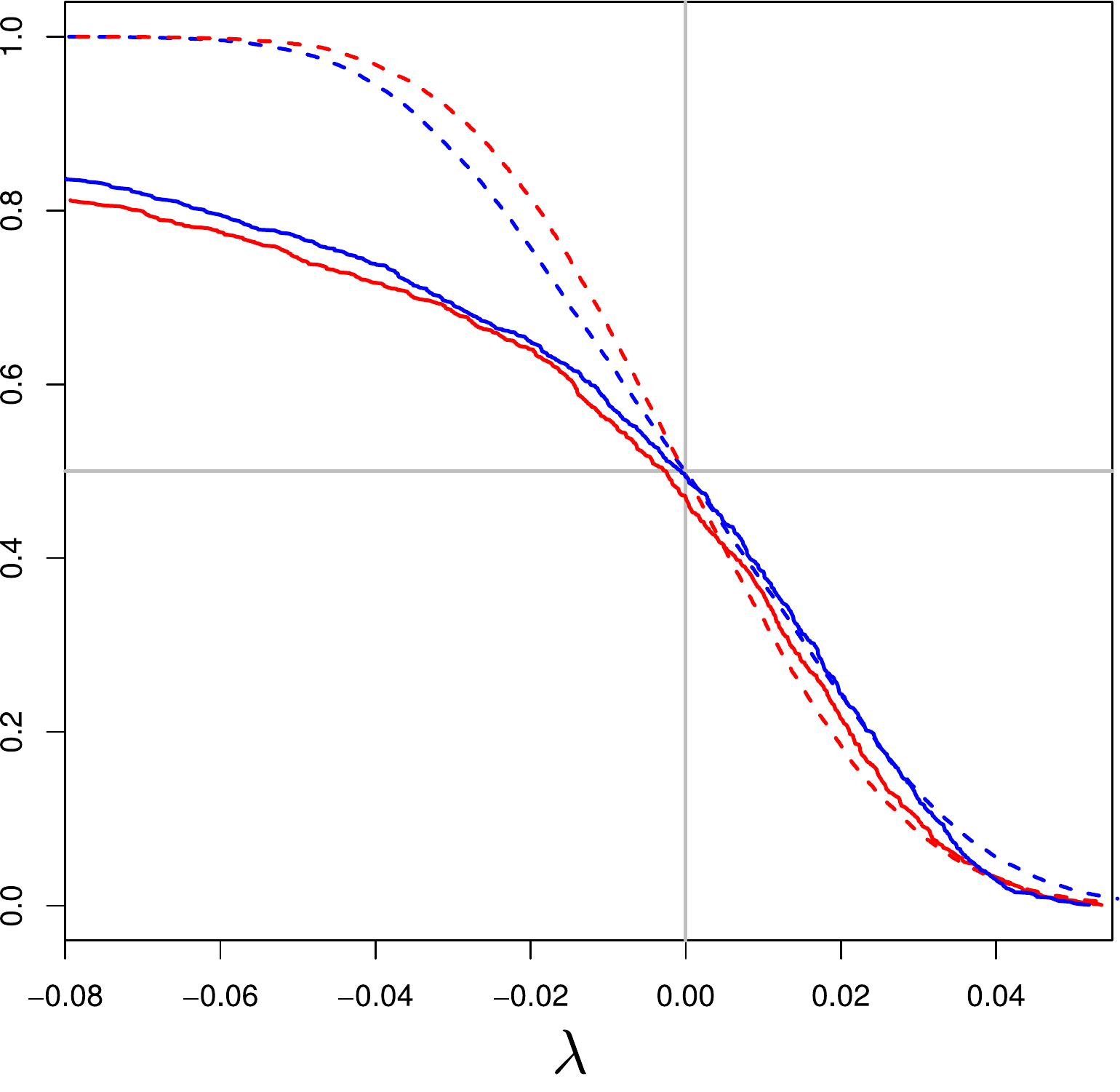}
   \caption{Cumulative distribution function for the chirality
     distribution of the $A_1$ ensemble. Red lines refer to $\nu=1$,
     blue lines to $\nu=2$. Solid lines are obtained from the
     QCD simulation, while dashed lines indicate the expectation
     from WRMT.}
   \label{fig:chirality_cdf}
\end{figure}

While real eigenmodes can and do appear for the topologically trivial
sector, a chirality distribution equal to zero is guaranteed because
of the symmetry between positive and negative values of the
topological charge. In order to facilitate a quantitative comparison
between the chirality distributions of QCD and the ones from WRMT in
the sectors with non-trivial topology we again consider the cumulative
distributions.  Figure~\ref{fig:chirality_cdf} shows those
found in ensemble $A_1$ from configurations with $\nu=1$ (red) and 2
(blue) in solid lines, with the WRMT predictions in dashed lines. Note
that in this plot, $m_\text{cut}$ would correspond to a value of
$\lambda=-0.3$.

We find a clear distinction between the QCD and the WRMT
distributions, as the QCD results are asymmetric around 0. It is
interesting, however, to observe that the distributions do in fact
match quite well up to the determined value of the critical
mass. Beyond that point, the chirality distribution as described by
WRMT drops to zero quite rapidly, while crossings keep occurring for
the QCD theory. The fact that the derivative of the absolute charge
with respect to the mass cut-off has become very small for larger
values of the cut-off (cf.~table \ref{tab:Nf0_simulation_parameters}) shows that this deviation
cannot be attributed straightforwardly to a lack of statistics or
misassigned configurations. On the other hand, the structure of the
WRMT prevents any of the operators included from breaking the
symmetry of the distribution around 0, so the mismatch must be due to
lattice artefacts in the chirality distributions which appear not to be
described by WRMT.

\subsubsection{Additional modes}
\label{subsec:additional}
\begin{figure}[!ht]
   \centering
   \includegraphics[width=0.7\textwidth]{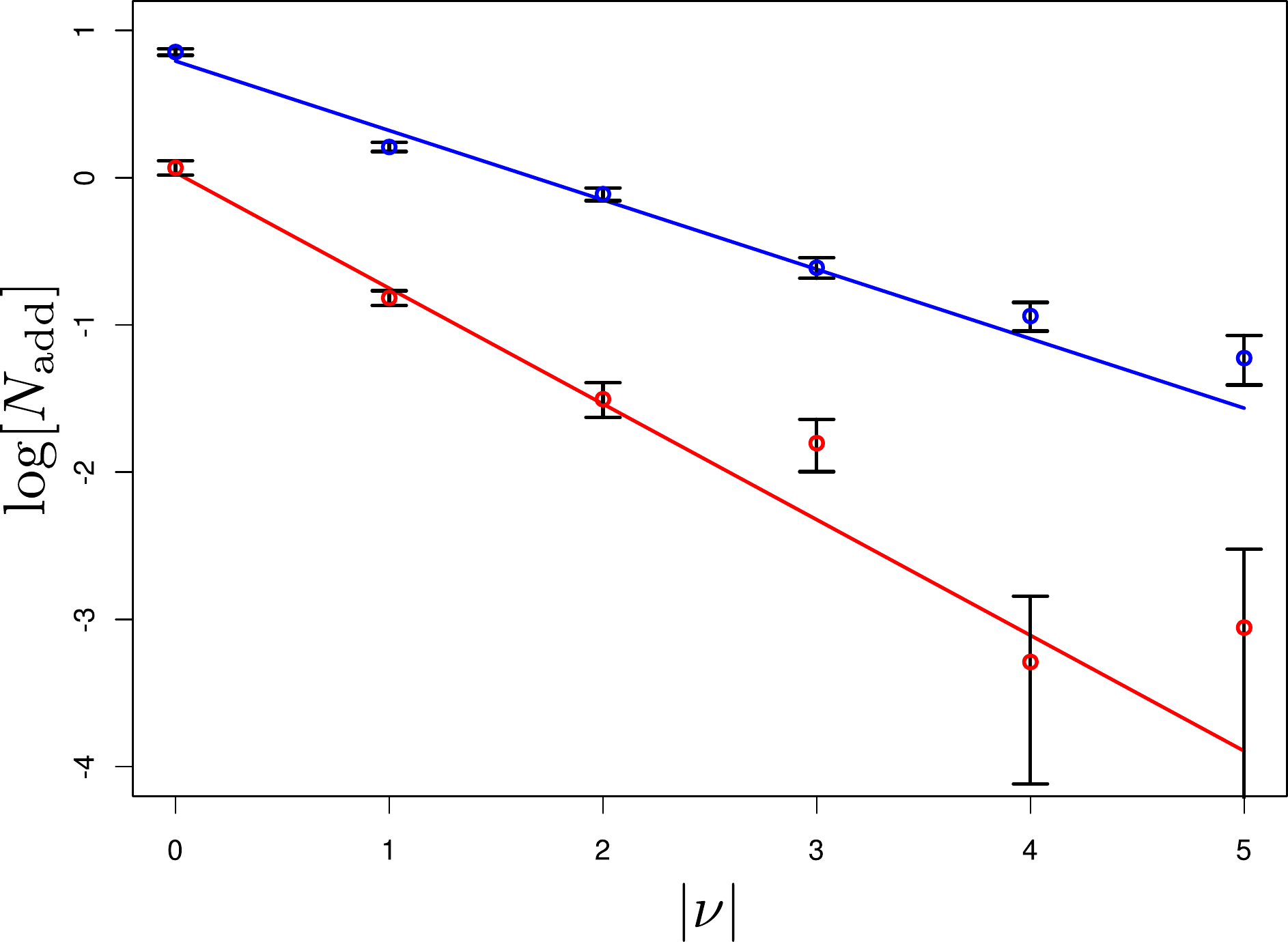}
   \caption{The logarithm of the average number of additional
     crossings $N_\mathrm{add}$ versus the absolute charge
     $|\nu|$. Shown are values for ensemble $A_1$ (red) and $B_1$ (blue).}
   \label{fig:ac_log}
\end{figure}

Apart from the chirality distribution, WRMT and WChiPT allow for
inferring a relationship between the charge sector and the average
number of additional crossings $N_\mathrm{add}$, i.e., the difference
between the total number of crossings found in the Wilson flow and the
absolute charge of the configuration. For small lattice spacing $a$,
this relationship is given in \cite{Kieburg:2011uf} as
\be
N_\mathrm{add} \propto a^{2(\nu + 1)}, 
\ee
a pure power law. 
The determination of $N_\mathrm{add}$ becomes progressively harder for
larger values of the charge, as both the number of such configurations
and the expectation value of $N_\mathrm{add}$ decrease rapidly.
Nevertheless, the data for ensembles A1 and B1 displayed in
figure~\ref{fig:ac_log} appears consistent with the predicted
power-law behaviour.  Moreover, the ratio of the coefficients of the
linear terms should be given by the ratio of the lattice spacings.
Performing linear fits to the initial three (for $A_1$) and four (for
$B_1$) points, we obtain the coefficients -0.785 and -0.471,
respectively.  The ratio then comes out as 1.67, to be compared to
1.64 from the ratio of lattice spacings as given by $r_0/a$. This
excellent agreement may be somewhat accidental here, given the limited
number of reliable points and the clear sensitivity to the fit
range. Nevertheless, we conclude that there are strong indications
that the power law for $N_\mathrm{add}$ derived
in~\cite{Kieburg:2011uf} holds for our data, in spite of the limited
agreement with the prediction for the chirality distribution and the
somewhat coarse lattice spacing in $B_1$.

\section{Conclusions}
We have investigated the spectral properties of the Wilson Dirac
operator on the lattice for $N_f=0$ quenched QCD in the
$\epsilon$-regime by numerical simulations. The resulting eigenvalue
distributions $\rho_5(\lambda)$ of the hermitian Wilson Dirac operator
$\D5$ and the chirality distributions $\rho_\chi(\lambda)$ of the real
modes of the Dirac operator $\DW$ can be compared to expressions
resulting from Wilson chiral perturbation theory (WChiPT) in the
$\epsilon$-regime, or equivalently by Wilson random matrix theory (WRMT) in
the microscopic limit. In order to do so, it is important to have good
control over the different topological sectors of QCD and we have
taken care of this by employing the topological index based on the
eigenvalue flow of $\D5$ as the definition of the topological
charge. Towards the continuum limit this definition becomes equivalent
to the index of the overlap operator.

Concerning the eigenvalue distributions $\rho_5^k(\lambda)$ of the
$k$-th eigenvalue, we find that they can be described rather well by
the corresponding distributions as given by the WRMT. It turns out
that by fitting these distributions, it is indeed possible to extract
the leading low energy constants $W_6, W_7$ and $W_8$ for Wilson
fermions from the eigenvalue spectra of the Wilson Dirac operator in
the $\epsilon$-regime, on volumes as small as $V \sim (1.5
\text{fm})^4$ and lattice spacings as coarse as $a \sim 0.1
\text{fm}$. However, in order to describe the shape of the
distributions properly, it is in fact necessary to include the effects
proportional to $W_6$ and $W_7$ in addition to the ones proportional
to $W_8$. From the resulting fits, one can extract values for $\Sigma
V$ at the one percent level.  However, since these values should
rather be regarded as effective parameters to match the spectral
distributions of $\D5$ with those of $\tilde D_5$, we refrain from
quoting a physical value for the corresponding quark condensate.

Our results indicate that the description of the microscopic Wilson
Dirac operator spectrum by WRMT works best in the topological charge
sectors with low values of the charge and starts to break down as the
topological charge increases, the more so the smaller the physical
volume. In addition, the chirality distributions 
obtained from our simulations show an asymmetry around 0 with an
enhanced tail for $\lambda_k^\text{W} > 0$ which can not be described
by the operators included in WRMT and WChiPT parametrising the leading
order lattice discretisation effects. It appears that the mixing of
eigenmodes of different chiralities due to the breaking of chiral
symmetry by the Wilson term is stronger than anticipated by the
description of WRMT and WChiPT, at least at the lattice spacings
investigated in this work. Nevertheless, an important result of WRMT
is that the number of additional real modes in excess of the
topological charge $\nu$ is suppressed for large values of $\nu$
\cite{Kieburg:2011uf} and we do find that this is indeed the case.

It would be particularly interesting to see how WRMT works for the
special case of $N_f=1$ QCD. In that situation there is no spontaneous
breaking of chiral symmetry, but a WRMT description is nevertheless
available. Since the theory defined with Wilson fermions suffers from
a sign problem towards the $\epsilon$-regime, any information on the
spectral properties would be useful in order to avoid numerical instabilities
of such simulations.

\subsection*{Acknowledgements}
We would like to thank Wolfgang Bietenholz and Tom DeGrand for inspiring remarks.

We learned at the Lattice 2011 meeting that Poul Damgaard, Urs Heller
and Kim Splittorff have independently been investigating the issues we
have discussed here. We refer the reader to their paper \cite{Damgaard:2011eg}.

\bibliography{wilson_Dirac_spectra}
\bibliographystyle{JHEP}
\end{document}